\documentclass[final,5p,times,twocolumn]{elsarticle}
\usepackage{graphics}
\usepackage{amsmath}
\usepackage{amssymb}
\usepackage{amsfonts}
\usepackage{graphicx}
\usepackage{subfigure}

\biboptions{numbers,sort&compress}

\usepackage{tabularx}
\usepackage{multirow}
\usepackage{booktabs}
\usepackage{ctable}
\newcolumntype{Y}{>{\raggedright\arraybackslash}X} 
\newcolumntype{W}{>{\raggedleft\arraybackslash}X}  
\newcolumntype{Z}{>{\centering\arraybackslash}X}   

\journal{Microelectronics Engineering}

\begin{document}

\begin{frontmatter}

\title{Energetics of native point defects in GaN: a density-functional study}

\author{Giacomo Miceli\corref{cor}}
\ead{giacomo.miceli@epfl.ch}
\author{Alfredo Pasquarello}
\address{Chaire de Simulation \`a l'Echelle Atomique (CSEA), %
 Ecole Polytechnique F\'ed\'erale de Lausanne (EPFL), %
 CH-1015 Lausanne, Switzerland}
\cortext[cor]{Corresponding author}

%
%
\begin{abstract}
We study the formation energies of native point defects in GaN through density-functional
theory. In our first-principles scheme, the band edges are positioned in accord with
hybrid density functional calculations, thus yielding a band-gap in agreement with experiment.
With respect to previous semilocal calculations, the calculated formation energies and 
charge transition levels are found to be significantly different in quantitative terms, 
while the overall qualitative trend remains similar. In Ga-rich conditions, the nitrogen 
vacancy corresponds to the most stable defect for all Fermi energies in the band gap, but  
its formation energy is too high to account for autodoping. Our calculations also 
indicate that the gallium vacancy does not play any compensating role in $n$-type GaN.
\end{abstract}


\end{frontmatter}

%
%
\section{Introduction}
\label{sec:intro}
The discovery of $p$ conductivity in magnesium-doped GaN and the achievement of
highly efficient blue-light-emitting diodes \cite{amano_JJAP1989,nakamura_JJAP1992,nakamura_APL1994}
have made of GaN one of the most promising compounds for future optoelectronic 
device applications in the visible and ultra-violet frequency range \cite{nakamura_IEEE2013}.
Furthermore, its wide and direct band gap, high thermal conductivity,
favorable electron transport, and large breakdown fields also make this compound ideally
suited for high-power and high-frequency electronic devices \cite{pearton_JAP1999}.
Despite the rapid commercialization of GaN-based devices, their performance is still not optimal. 
For instance, while the current $p$-doping level is sufficient for light-emitting diodes, the lack
of an efficient $p$-doping is recognized as the bottleneck for obtaining GaN-based
laser diodes. Native defects are commonly thought to play a role in the limited 
hole conductivity of $p$-type GaN. For some time, native defects such as 
nitrogen vacancies have also been invoked for explaining the $n$-type conductivity of 
as-grown nonintentionally doped GaN samples \cite{maruska_APL1969,ilegems_JPCS1973}.
However, the current interpretation is that this autodoping rather results from 
extrinsic impurities like oxygen, due to the high calculated defect formation 
energies of vacancy defects \cite{van2004first}. Native defects are also thought
to be responsible for shortening the lifetime of electronic devices and for their low
performances. In this context, it is clear that the availability of accurate  
calculations of native defects plays a critical role for further progress. 

Several density-functional studies have provided a comprehensive description of
the structural and energetic properties of native point defects in GaN
\cite{neugebauer_PRB1994,boguslawski_PRB1995,neugebauer_MRSP1995,limp_PRB2004}.
However, these studies generally relied on semilocal density functionals which severely
underestimate the band gap of GaN. Furthermore, the common practice of aligning the calculated 
valence band edge to the experimental one further led to incorrectly positioned defect 
levels. Early calculations also suffered from small simulation cells and inappropriate 
treatment of the finite-size effect. Given their potential role in many of the observed 
electronic phenomena of GaN, it is highly desirable to revisit the formation energies 
of native defects in GaN through the use of state-of-the-art methodologies.

In this work, we address the formation energies of native point defects in GaN using 
a first-principles scheme which reproduces the band gap of GaN and overcomes the alignment
problem. Vacancies, interstitials, and antisites are studied in all their stable charge 
states. We find significantly different formation energies and defect levels than 
in previous calculations, but the general qualitative trends are confirmed. 
In Ga-rich conditions, the nitrogen vacancy is found to be the most stable defect 
throughout the band gap, but its formation energy is too high to account for autodoping. 
Our calculated formation energies indicate that the gallium vacancy is unlikely
to play any compensating role in $n$-type conditions. 

%
%
\section{Methods}
\label{sec:compdetails}

In this work, the electronic structure is described through the density functional  
proposed by Perdew, Becke, and Ernzerhof (PBE) \cite{pbe}. 
We use a computational framework based on norm-conserving pseudopotentials 
and plane-wave basis sets, as implemented in the {\sc q}uantum-\textsc{espresso} suite 
of programs \cite{quantum_espresso}. Gallium $3d$ electrons are explicitly included 
in the valence shell. The kinetic energy cut-off is set at 75 Ry. Spin-unrestricted 
calculations are performed when unpaired electrons occur. Only the lowest spin 
multiplicity are considered. All the defect structures
are fully relaxed. The Brillouin zone sampling is carried out through the use of an
off-center single $k$-point as defined in the Monkhorst-Pack scheme \cite{monkhorst_PRB1976},
corresponding to the equivalent of the Baldereschi point for an hexagonal lattice 
\cite{baldereschi}.

The formation energy of a defect $X$ in the charged state $q$ as a function of 
the Fermi energy $\epsilon_{\textrm{F}}$, referenced to the valence band maximum 
$\epsilon_\textrm{v}$ reads
\begin{eqnarray}\label{eq:formenerg}
E_{\textrm{f}}[X^q] &=& E_{\textrm{tot}}[X^q] - E_{\textrm{tot}}[\textrm{bulk}] 
                      - \sum_{\alpha}{n_{\alpha} \mu_{\alpha}} + \nonumber\\ && 
                      + q (\epsilon_{\textrm{F}}+\epsilon_\textrm{v}) 
                      + E^q_{\textrm{corr}}
\end{eqnarray}
where $E_{\textrm{tot}}[X^q]$ and $E_{\textrm{tot}}[\textrm{bulk}]$ are the total
energies of supercell calculations with and without the defect, respectively.
In Eq.\ (\ref{eq:formenerg}), $\mu_{\alpha}$ is the chemical potential of the atomic 
species $\alpha$ and $n_{\alpha}$ the number of atomic species $\alpha$ 
added ($n_{\alpha} > 0$) or removed ($n_{\alpha} < 0$) from the pristine compound. 
We consider Ga-rich and N-rich conditions, in which 
$\mu_\textrm{Ga}=\mu_\textrm{Ga}[\textrm{bulk}]$ and 
$\mu_\textrm{N}=\mu_{\textrm{N}_2}$, respectively. In either of the considered conditions,
the chemical potential of the other species is determined according to the thermodynamic
equilibrium condition in GaN: $\mu_\textrm{GaN} = \mu_\textrm{Ga} + \mu_\textrm{N}$.
$E^q_{\textrm{corr}}$ accounts for finite-size corrections
and includes a potential alignment term \cite{freysoldt_PRL2009,komsa_PRB2012}.

To overcome the band-gap underestimation problem of semilocal functionals, we use
the band-edge positions as found with the Heyd-Scuseria-Ernzerhof hybrid 
density functional (HSE) \cite{hse,hse_erratum,komsa_PRB2010}. We fixed the 
fraction of Fock exchange to $\alpha=0.31$ in order to reproduce the experimental 
band gap of GaN. Through an alignment based on the average electrostatic potential,
the band-edge positions achieved with the HSE functional are then
positioned with respect to the electronic structure achieved in the PBE 
\cite{dahinden,miceli_ME2013}. The validity of this alignment scheme rests 
on the fact that energy levels of atomically localized defects do not vary 
when going from the PBE to the hybrid-functional description 
\cite{AlkauskasPRL2008,PSSBAudrius,ZnOAudrius,komsa_PRB2010}.
For defect states involving anionic 2$p$ orbitals, self-interaction 
effects leading to charge localization are better described in hybrid-functional 
approaches \cite{pacchioni_PRB2001,laegsgaard_PRL2001}. However, these effects
generally do not affect the defect formation energy in a significicant way 
\cite{carvalho_PRB2009}.

%
%
\section{Results}
\label{sec:results}

\begin{figure}
\centering
\includegraphics[width=1.0\columnwidth]{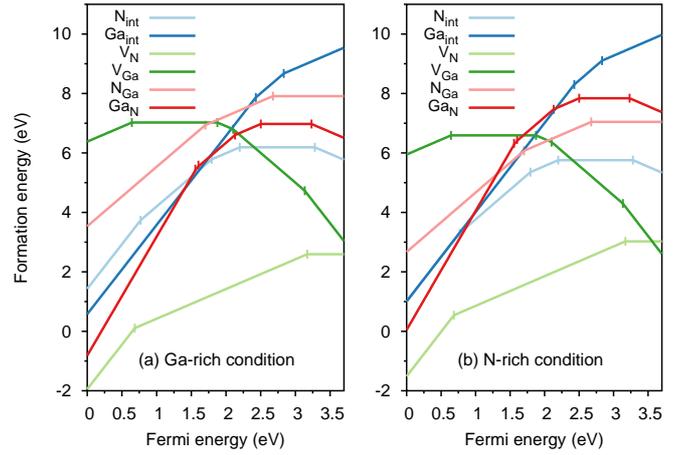}
\caption{Calculated formation energies of native point defects in GaN as a function
         of Fermi energy. In panels (a) and (b), Ga-rich and N-rich conditions are
         assumed, respectively. For each defect, only the lowest formation energy
         is shown for varying Fermi energy. The kinks in the solid line are marked
         by vertical bars and correspond to thermodynamic charge transition levels.}
\label{fig:energies}
\end{figure}

All the defects are modeled starting from a 108-atom supercell of wurtzite GaN 
subject to periodic boundary conditions. The larger simulation cell used here 
allows for a better description of the atomic relaxation around the defects compared
to earlier studies \cite{neugebauer_PRB1994,boguslawski_PRB1995}. 
The calculated formation energies of all native defects in GaN
are shown as a function of the Fermi energy in Fig.\ \ref{fig:energies}, 
for both Ga-rich and N-rich conditions. The formation energies achieved 
for the Fermi energy at the top of the valence band are reported in 
Table \ref{tab:energies}. The corresponding charge transition levels 
are summarized in Table \ref{tab:ctl}.  Compared to previous semilocal calculations 
\cite{boguslawski_PRB1995,van_JAP2004,limp_PRB2004},
we notice significant quantitative differences reaching up to 2 eV for formation
energies and up to 1 eV for charge transition levels. Nevertheless, the overall 
qualitative trends reported in Refs.\ \cite{van_JAP2004,limp_PRB2004} can still 
be recognized. Variations can be attributed to various effects, including the 
size of the simulation cell, the {\bf k}-point sampling, the treatment of Ga $3d$
states, the finite-size corrections, and the alignment procedure.

\begin{table}
 \caption{Formation energies of native point defects in GaN calculated for the Fermi
          level at the top of the valence band in Ga-rich and N-rich conditions.}
 \label{tab:energies}
  \centering
  \begin{tabular}{lccc}
  \hline \hline
                            & & \multicolumn{2}{c}{E$_{\textrm{f}}$ (eV)} \\
                            \cline{3-4}\noalign{\smallskip}
  Defect                    & Charge state & Ga-rich & N-rich \\
  \hline
  N$_{\textrm{int}}$        & +3           &  1.43 & 0.99 \\
                            & +2           &  2.20 & 1.76 \\
                            & +1           &  4.00 & 3.56\\
                            & 0            &  6.19 & 5.75\\
                            & $-1$         &  9.47 & 9.03\\
  &&&\\
  Ga$_{\textrm{int}}$       & +3           &  0.58 & 1.02 \\
                            & +2           &  3.01 & 3.45 \\
                            & +1           &  5.84 & 6.28 \\
  &&&\\
  V$_{\textrm{N}}$          & +3           & $-1.95$ & $-1.51$ \\
                            & +1           & $-0.58$ & $-0.14$ \\
                            & 0            &  2.59   & 3.03 \\
  &&&\\
  V$_{\textrm{Ga}}$         & 0            &  7.02 & 6.58 \\
                            & $-1$         &  8.90 & 8.46 \\
                            & $-2$         & 11.00 & 10.56 \\
                            & $-3$         & 14.14 & 13.70 \\ 
  &&&\\
  N$_\textrm{Ga}$           & +2           &  3.53 & 2.66 \\
                            & +1           &  5.23 & 4.36 \\
                            & 0            &  7.91 & 7.04 \\
  &&&\\
  Ga$_\textrm{N}$           & +4           & -0.82 & 0.05 \\
                            & +3           &  0.79 & 1.66 \\
                            & +2           &  2.34 & 3.21 \\
                            & +1           &  4.47 & 5.34 \\
                            & 0            &  6.97 & 7.84 \\
                            & $-1$         & 10.21 & 11.08 \\
  \hline \hline
  \end{tabular}
 
\end{table}

\begin{table}
 \caption{Calculated charge transition levels $\epsilon_{q/q'}$ between the charges states 
          $q$ and $q'$ for native point defects in GaN. The defect levels are referred to 
          the top of the valence band.}
 \label{tab:ctl}
  \centering
  \begin{tabular}{lcc}
  \hline \hline
  Defect                    & $q/q'$ & $\epsilon_{q/q'}$ (eV) \\
  \hline
  N$_{\textrm{int}}$        & +3/+2        & 0.77 \\
                            & +2/+1        & 1.80 \\
                            & +1/0         & 2.20 \\
                            & 0/$-1$       & 3.28 \\
  &&\\
  Ga$_{\textrm{int}}$       & +3/+2        & 2.43 \\
                            & +2/+1        & 2.83 \\
  &&\\
  V$_{\textrm{N}}$          & +2/+1        & 0.68 \\
                            & +1/0         & 3.17 \\
  &&\\
  V$_{\textrm{Ga}}$         & 0/$-1$       & 1.88 \\
                            & $-1$/$-2$    & 2.10 \\
                            & $-2$/$-3$    & 3.13 \\ 
  &&\\
  N$_\textrm{Ga}$           & +2/+1        & 1.70 \\
                            & +1/0         & 2.67 \\
  &&\\
  Ga$_\textrm{N}$           & +4/+3        & 1.56 \\
                            & +3/+2        & 1.60 \\
                            & +2/+1        & 2.13 \\
                            & +1/0         & 2.50 \\
                            & 0/$-1$       & 3.23 \\

  \hline \hline
  \end{tabular}
\end{table}

\begin{figure}
\centering
\includegraphics[width=1.0\columnwidth]{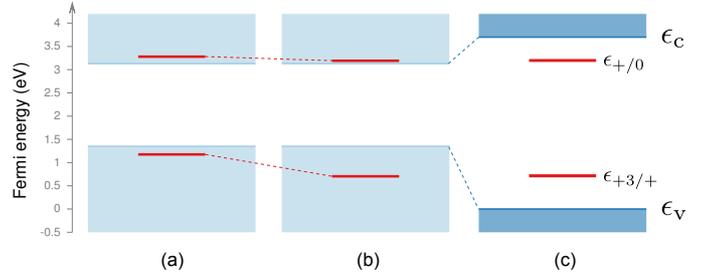}
\caption{Charge transition levels of the nitrogen vacancy calculated with the PBE functional
         (a) without and (b) with finite-size corrections. In panel (c) the 
         correct band gap is obtained through the use of the HSE hybrid functional. 
         The defect levels calculated at the PBE level are positioned in the HSE 
         band-gap by aligning the electronic structures through the average 
         electrostatic potentials.}
\label{fig:ctl}
\end{figure}

It should be stressed that a pure semilocal calculation may yield
quantitatively and qualitatively erroneous results. To illustrate the shifts undergone 
by the defect energy level upon the variation of the adopted level of theory, 
we focus on the nitrogen vacancy defect. This defect shows two charge transition
levels in the band gap, $\epsilon_{+3/+}$ and $\epsilon_{+/0}$, as
shown in Fig.\ \ref{fig:energies}. In a pure PBE calculation, the calculated band gap
of 1.78 eV severely underestimates the experimental one (3.5 eV, Ref.\ \cite{madelung2004})
and neither of the two charge transition levels are found in the band gap, 
as shown in Fig.\ \ref{fig:ctl}(a). We can nevertheless
obtain the position of the defect levels by sampling the Brillouin zone 
through a single off-center {\bf k}-point. In such a setup, the energy gap in the calculation 
increases to 3.95 eV, the defect charge distributions are atomically well localized, 
and the two charge transition levels can meaningfully be determined.  
The obtained defect energy levels can then be positioned with respect to the 
actual GaN band edges obtained from a bulk calculation with a dense {\bf k}-point 
sampling through the alignment of the average electrostatic potential.
Next, finite-size effects need to be considered as they are still significant for 
a simulation cell of 108 atoms. For instance, for the nitrogen vacancy in its +3 
charge state, the correction term due to spurious electrostatic interactions
is as large as 1.04 eV. For the Ga$_\textrm{N}$ defect in its +4 charge state,
the finite-size correction reaches 2.12 eV, due to the high nominal charge of the defect.
The effect of the finite-size corrections on the charge transition levels of the 
nitrogen vacancy are illustrated in Fig.\ \ref{fig:ctl}(b). In the final step of our 
procedure, the band edges obtained with the HSE functional are positioned
on the same energy scale as the defect levels obtained with the PBE functional.
This is achieved through the alignment of the bulk electrostatic potentials 
in the HSE and PBE calculations \cite{AlkauskasPRL2008,PSSBAudrius,ZnOAudrius,komsa_PRB2010}. 
As shown in Fig.\ \ref{fig:ctl}(c), this last step produces the opening of 
the band gap which thereby encloses the two charge transition levels of the 
nitrogen vacancy. 

The validity of our methodological scheme can further be assessed
by comparing our results for the nitrogen vacancy with those obtained within 
a fully consistent hybrid functional approach \cite{yan_APL2012}. The +3/+ and 
+/0 charge transition levels found in the present work at 0.68 and
3.17 eV compare very well with the respective hybrid-functional levels at 0.47 
and 3.25 eV \cite{yan_APL2012}. Similarly, the formation energy 
of the neutral state found at 2.6 eV in our calculation (cf.\ Ga-rich 
conditions in Fig.\ \ref{fig:energies}) shows a fair agreement with the 
corresponding hybrid functional result of 3.1 eV \cite{yan_APL2012}. 
The quality of this agreement provides confidence in the accuracy of 
our calculation scheme. Remarkably, this level of accuracy is
presently achieved with PBE calculations which imply a significantly lower
computational cost than fully consistent hybrid functional calculations,
thereby enabling us to comprehensively address the energetics of native 
defects in GaN.

In Ga-rich conditions, the nitrogen vacancy is found to be the most
stable defect for all Fermi energies in the band gap. In particular, this condition 
also applies for Fermi energies close to the conduction band, unlike 
previous calculations which found the gallium vacancy at lower energies \cite{limp_PRB2004}. 
In $p$-type conditions, the calculated formation energy of the nitrogen vacancy 
is $-$2.0 eV, indicating that such defects would lead to strong compensation 
effects in this regime. In $n$-type conditions, the formation energy is 2.6 eV, 
lower by more than 1 eV with respect to previous semilocal calculations \cite{limp_PRB2004}, 
but still too high to assign a role to this defect in the observed $n$-type 
conductivity of nonintentionally doped GaN. Indeed, the calculated formation
energy of the nitrogen vacancy in its charge state +1 results in an $n$-type 
carrier concentration of $\sim 10^{11}$ cm$^{-3}$ at 
typical growth temperatures around 1050$^\circ$C \cite{ilegems_JPCS1973}.
This level of autodoping is orders of magnitude lower than the lowest $n$-type 
carrier concentration measured in nominally undoped samples,
which ranges from $10^{17}$ to $10^{19}$ cm$^{-3}$ \cite{ilegems_JPCS1973}.
This further supports the conclusion drawn from previous calculations 
that the observed $n$-type carriers do not result from intrinsic defects 
such as nitrogen vacancies \cite{neugebauer_PRB1994,boguslawski_PRB1995}.  

In N-rich conditions, the nitrogen vacancy is still the lowest energy defect for 
all Fermi energy in the band gap but a small energy range in the vicinity of 
the conduction band, where the gallium vacancy in its $-$3 charge state shows 
slightly lower formation energies. However, the calculated formation energies 
remain too high for the gallium vacancy playing any relevant compensating 
role in $n$-type conditions.

%
%
\section{Conclusion}
\label{sec:conclusion}
We provide a comprehensive and accurate description of formation energies 
of native point defects in GaN using a state-of-the-art first-principles 
methodology. Such a description is fundamental for the interpretation of 
the electronic properties of GaN related to such defects.

\section*{Acknowledgements}
Financial support is acknowledged from the Swiss National Science
Foundation (Grants Nos.\ 200020-152799 and 200020-134600).
We used computational resources of CSCS and CSEA.

\end{document}